\documentclass[conference]{IEEEtran}
\IEEEoverridecommandlockouts
\usepackage{cite}
\usepackage{amsmath,amssymb,amsfonts}
\usepackage{algorithmic}
\usepackage{graphicx}
\usepackage{textcomp}                       
\usepackage{xcolor}
\usepackage{booktabs}
\usepackage{subcaption}
\usepackage{tcolorbox}
\usepackage{multirow}
\usepackage{url}

\def\BibTeX{{\rm B\kern-.05em{\sc i\kern-.025em b}\kern-.08em
    T\kern-.1667em\lower.7ex\hbox{E}\kern-.125emX}}

\begin{document}

\title{Contemporary COBOL: Developers' Perspectives on Defects and Defect Location 
}

\iftrue
\author{\IEEEauthorblockN{Agnieszka Ciborowska}
\IEEEauthorblockA{
\textit{Virginia Commonwealth University}\\
Richmond, Virginia, USA\\
ciborowskaa@vcu.edu}
\and
\IEEEauthorblockN{Aleksandar Chakarov}
\IEEEauthorblockA{
\textit{Phase Change Software}\\
Golden, Colorado, USA\\
achakarov@phasechange.ai}
\and
\IEEEauthorblockN{Rahul Pandita}
\IEEEauthorblockA{
\textit{Phase Change Software}\\
Golden, Colorado, USA\\
rpandita@phasechange.ai}
}
\fi
\maketitle

\begin{abstract}
Mainframe systems are facing a critical shortage of developer workforce as the current generation of COBOL developers retires. Furthermore, due to the limited availability of public COBOL resources, entry-level developers, who assume the mantle of legacy COBOL systems maintainers, face significant difficulties during routine maintenance tasks, such as code comprehension and defect location. While we made substantial advances in the field of software maintenance for modern programming languages yearly, mainframe maintenance has received \textit{limited} attention. With this study, we aim to direct the attention of researchers and practitioners towards investigating and addressing challenges associated with mainframe development. Specifically, we explore the scope of defects affecting COBOL systems and defect location strategies commonly followed by COBOL developers and compare them with the modern programming language counterparts. To this end, we surveyed 30 COBOL and 74 modern Programming Language (PL) developers to understand the differences in defects and defect location strategies employed by the two groups. Our preliminary results show that: (1) major defect categories affecting the COBOL ecosystem are different than defects encountered in modern PL software projects; (2) the most challenging defect types in COBOL are also the ones that occur most frequently; and (3) COBOL and modern PL developers follow similar strategies to locate defective code.

\end{abstract}


\section{Introduction}
\label{sec:intro}

Mainframe systems, although perceived by many as technologically outdated, stand at the core of the daily operations of many financial, health care, and governmental institutions.
They provide business-essential features that enable rapid, reliable, and secure processing of extremely large volumes of transactions.
To put the scale and importance of the mainframe development into perspective, 
there are \textit{``over 220 billion lines of COBOL code being used in production today, and 1.5 billion lines are written every year''}~\cite{forbes-cobol},
while \textit{``on a daily basis, COBOL systems handle USD 3 trillion in commerce''}~\cite{newstack-cobol}.

These staggering numbers emphasize two important facts. First, mainframe development is resilient and unlikely to be replaced any time soon due to significant cost and risk associated with building/migrating to new transaction processing systems that realize existing, well-established functionalities. 
Second, as this generation of mainframe developers retires, support for these systems is in jeopardy as mainframe development is not part of current mainstream curricula~\cite{carr2000cobol,kizior2000does}.
%Anecdotally, US Stimulus checks needed to financially assist Americans adversely affected by the COVID-19 situation, were delayed due to limited understanding of the COBOL system~\cite{covid-bloomberg}.
We suspect this problem will only worsen as the new generation of developers trained on modern programming stacks takes the responsibility of maintaining legacy COBOL systems. 

While research efforts in studying maintenance in modern PL stacks resulted in advanced tooling~\cite{sando,martinez2019rtj,mai2020smrl,buhse2019vedebug,lockwood2019mockingbird,meng2019convul,beyer2019testcov} and guidance to improve developer productivity~\cite{fritz2016leveraging, goncales2019measuring}, investigating maintenance of mainframe systems is rare with most studies focusing on language migration~\cite{sneed2001,demarco2018cobol,rodriguez2013bottom}.
Given the low success rate (less than 50\%\cite{tech-republic}) of the mainframe migration effort, we posit that mainframe developers need support and innovation in the most routine software maintenance activities, such as locating and fixing software defects~\cite{latoza2006}.

%While developers working with modern programming languages have seen a multi-fold increase in developer tooling~\cite{sando,martinez2019rtj,mai2020smrl,buhse2019vedebug,lockwood2019mockingbird,meng2019convul,beyer2019testcov}, tools assisting mainframe developers have remained fairly stagnant.
%This leads to vastly different experiences when performing even the most common software development activities, such as locating and fixing software defects~\cite{latoza2006}. 
%While significant progress has been made tackling software defects, the development and evaluation of such tools and approaches are frequently limited to open-source software projects leveraging modern technologies.
%As a result, the difficulty of applying these tools in different development settings and the scale of required adjustments remain unknown.

As a step towards better understanding the needs of mainframe developers, \textit{in this study, we explore mainframe developers’ perspectives regarding the scope of software defects affecting mainframe systems and the strategies commonly followed to locate the defects in the legacy codebase.} We further investigate and highlight the differences in defects and defect location strategies as reported from the perspectives of COBOL and non-COBOL developers. Through this comparison, we aim to: (1) provide insights into the types and frequency of defects encountered during the development and maintenance of mainframe COBOL systems; and (2) elicit a key features of a hypothetical defect location tool targeted towards supporting mainframe developers. In particular, this work addresses the following research questions:

\noindent\textbf{RQ1: What are the major categories of software defects in COBOL? Are these defect types specific to COBOL?}
Previous research identified different defect categories and analyzed how frequently these defects occur in code~\cite{odc,catolino2019,lal2012,bruning2007}. However, most of the analyzed software ecosystems are built around the object-oriented paradigm, leaving a gap in understanding whether the same problems affect other environments. This study focuses on mainframe projects written in COBOL and contrasts frequent defects reported by COBOL and modern PL developers.

\noindent\textbf{RQ2: Are challenging software defects the same as typical defects? What are the major challenging software defects in COBOL and non-COBOL environments?}
Little is known about the developers' point of view on the defects that are most challenging to locate and fix~\cite{kocchar2016,siegmund2014}. The goal of RQ2 is to contrast the types and properties of typical and challenging software defects. This comparison can shed light on the less studied area of challenging software defects and pinpoint specific defect types that mainframe developers need the most assistance with.

\noindent\textbf{RQ3: Does the defect location process vary between software ecosystems? What are the similarities and differences in developer approaches to locating defective code?} Locating a software defect is a time- and effort-consuming task. To gain more insights into how developers perform defect location, researchers performed multiple lab and field studies to observe developers' work patterns~\cite{kevic2017,damevski2016,wang2011,wang2013,siegmund2014}.
In this study, we investigate developers' perceptions of the most effective approaches to locate defects in their respective environments to observe how working in a specific environment (e.g., COBOL or non-COBOL) affects the defect location process.

To answer these questions, we asked software developers about their perspective on software defects and strategies to localize faulty code. 
We surveyed 106 professional developers with different job seniority, specializing in various technologies, including 30 COBOL developers. To support the survey's findings and get more insight into developer viewpoints, we conducted 10 follow-up interviews.

The contributions of this paper are:
\begin{itemize}
    \item identification of the most frequent and challenging software defects in COBOL,
    \item investigation of developers' perception of the most useful defect location strategies,
    \item recommendations of the key features of a hypothetical defect location tool to support developers in maintaining mainframe COBOL systems,
    \item publicly available anonymized study materials to aid replication and foster future research in this area\cite{projectWeb}.
\end{itemize}

\section{Study design}
\label{sec:studyDesign}
\begin{figure} [ht]
	\centering
	\includegraphics[trim=100 80 90 80,clip,width=0.9\linewidth]{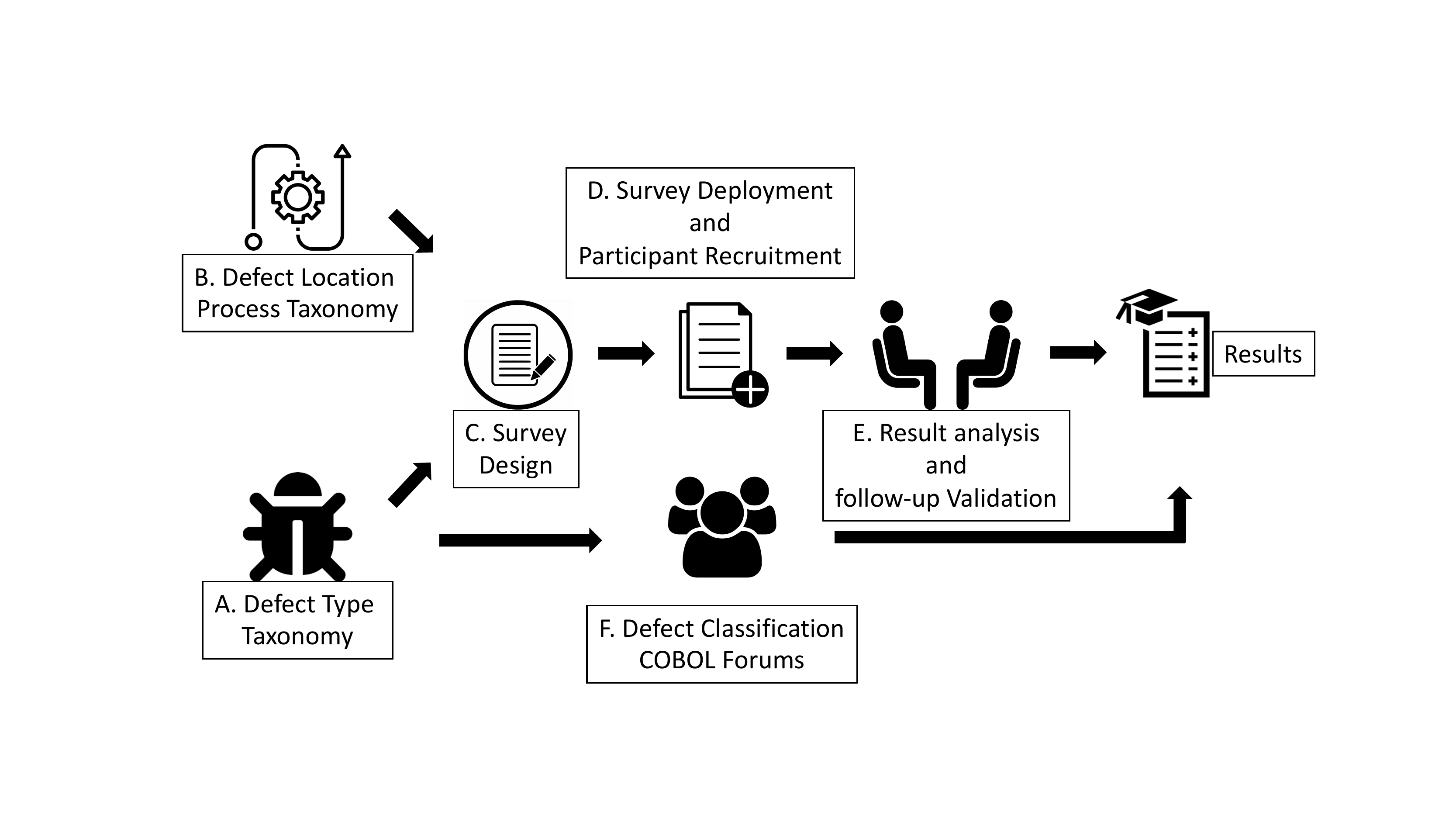}
	\caption{\small{An overview of our study design methodology.}}
	\label{fig:approach}
	 \vspace{-6mm}
\end{figure}

For this study, we employed an online survey due to the following reasons. First, the population of COBOL developers is relatively small and scattered, hence to recruit a sufficient number of participants we could not be limited by our geographical area. Second, COBOL projects are of crucial value to companies, so accessing their internal data was highly unlikely. Third, due to the rapidly evolving COVID-19 pandemic, an online study remained the safest choice.
We employed a mixed-methods approach to ascertain the differences between the COBOL and non-COBOL developer populations regarding defect types and defect location strategies. 
Specifically, we underwent the following stages, as depicted in Figure~\ref{fig:approach}:
(A) Defect types taxonomy synthesis, (B) Defect location strategy taxonomy synthesis, (C) Survey Design, (D) Survey Deployment and Participant Recruitment, (E) Result analysis and follow-up validation, and (F) COBOL Forums Defect Classification. We next describe each of these stages in detail.

\subsection{Defect Type Taxonomy}
\label{studyDesign_defectTaxonomy}
Previous research has identified various taxonomies characterizing software defects~\cite{odc,catolino2019,freimut2005,grady1992}. 
In this study, we leverage the classification schema proposed by Catolino et al.~\cite{catolino2019}, which provides a concise, yet comprehensive selection of defect types based on 9 categories: Configuration, Database, GUI, Network, Performance, Permission/Deprecation, Program Anomaly, Security, and Test Code. However, we note two major drawbacks of applying this taxonomy in our study.
First, Program Anomaly is a broad category and pertains to the majority of software defects.
This concern was identified during our think-aloud mock survey sessions (details in Sec.~\ref{sec:survey-design})(also mentioned  Catolino et al..~\cite{catolino2019}). 
Therefore, we decided to further expand Program Anomaly into 3 categories based on defect types indicated during the mock surveys. 
\textit{Input data} refers to issues caused by unexpected or malformed input data, which often occurs in the COBOL transaction processing system.
\textit{Concurrency-related} defects involve problems caused by e.g., incorrect access to shared resources or deadlocks. In contrast, \textit{Logic/Flow} defects pertain to programming errors, e.g., incorrect use of data structures. Second, participants of the mock survey session additionally identified a separate defect category related to issues arising due to the system operating at the limit of the resources, \textit{Workload/Stress}. In total, we use the 11 defect categories shown in Table~\ref{tab:defects-taxonomy}.

To capture additional software defect properties, we leverage Orthogonal Defect Classification (ODC)~\cite{odc}. ODC captures defect properties from two perspectives, when the defect is reported and when it is closed. 
Although ODC captures fine-grained details of defects, it is not applicable in the context of a survey since that requires participants to be familiar with extensive ODC terminology. Therefore, we opted for a hybrid approach where we leverage Catolino et al.~\cite{catolino2019} defect categories, which to some extent cover ODC's defect reporting perspective, to which we add three of the ODC's defect closing perspective: \textit{Age, Source}, and \textit{Qualifier}.
In particular, \textit{Age} refers to the time when the defect was introduced into the code base, \textit{Source} indicates which part of code was affected, whereas \textit{Qualifier} identifies the cause of the defect. Comparing the options specified in ODC for each of these dimensions, we decided to introduce two changes. First, we introduce \textit{Corrupted data} as a qualifier to capture one of the frequent causes of defects in COBOL projects. Second, we merge \textit{Outsourced} and \textit{Ported} categories in the \textit{Source} dimension into one category, \textit{Outsourced}. The modification was dictated 
by observing how during the mock survey session, the participants struggled to differentiate between the two options.
Finally, following Morrison et al.~\cite{morrison2018}, we introduced a dimension to determine an entity that reported a defect, \textit{Reporter}. The final version of defect location taxonomy is presented in Table~\ref{tab:defects-taxonomy} and includes the defects categories and their properties.

\begin{table}[ht]
\scriptsize
\caption{\small{Our software defect taxonomy based on \cite{odc} and \cite{catolino2019}.}}
\label{tab:defects-taxonomy}
\begin{tabular}{l|ll}
\toprule
\textbf{Defect categories} & \multicolumn{2}{l}{\textbf{Defect properties}} \\ \midrule
Concurrency & \multirow{3}{*}{Reporter} & Developer \\
Configuration &  & Quality assurance personnel/Tester \\
Database &  & End user \\ \cline{2-3}
GUI & \multirow{4}{*}{Cause} & Not coded to specification \\
Input Data &  & Missing or incomplete specification \\
Logic/Flow &  & Corrupted data \\ 
Network &  & Extraneous \\ \cline{2-3}
Performance & \multirow{3}{*}{Source} & Developed in-house \\
Permission/Deprecation &  & External library/API \\
Security &  & Outsourced \\ \cline{2-3}
Test Code & \multirow{4}{*}{Age} & New feature \\
Workload/Stress &  & Legacy code \\
 &  & Re-factored \\
 &  & Changed to fix a defect \\ \bottomrule
\end{tabular}
\
\end{table}

\subsection{Defect Location Taxonomy}
\label{studyDesign_defectLocationTaxonomy}

There is a rich history of exploratory observational studies investigating the process of defect location~\cite{ko2006,sillito2008,wang2013,kevic2017,damevski2016}.
In general, we note the following common strategies emerge across different studies: looking for initial code entities, exploring related program elements, and documenting relevant code components~\cite{ko2006,wang2013,sillito2008}.
In this study, we decided to leverage the hierarchical feature location model proposed by Wang et al.~\cite{wang2013}. 
The model comprises three levels of granularity: phases, patterns, and actions. Each phase represents a high-level stage in a feature location process, namely, \textit{Search} for entrance points, \textit{Expand} the search, \textit{Validate}, and \textit{Document}. Patterns define common strategies developers follow at different phases related to, e.g., execution or textual search, while actions correspond to fine-grained physical or mental activities undertaken by developers. 
Across all phases, the authors identified 11 physical actions and 6 mental actions. The physical actions included, e.g., reading code, or exploring source code files, while mental actions covered, e.g., conceiving an execution scenario or identifying keywords.

\begin{figure*}[ht]
    \centering
    \includegraphics[width=0.8\textwidth]{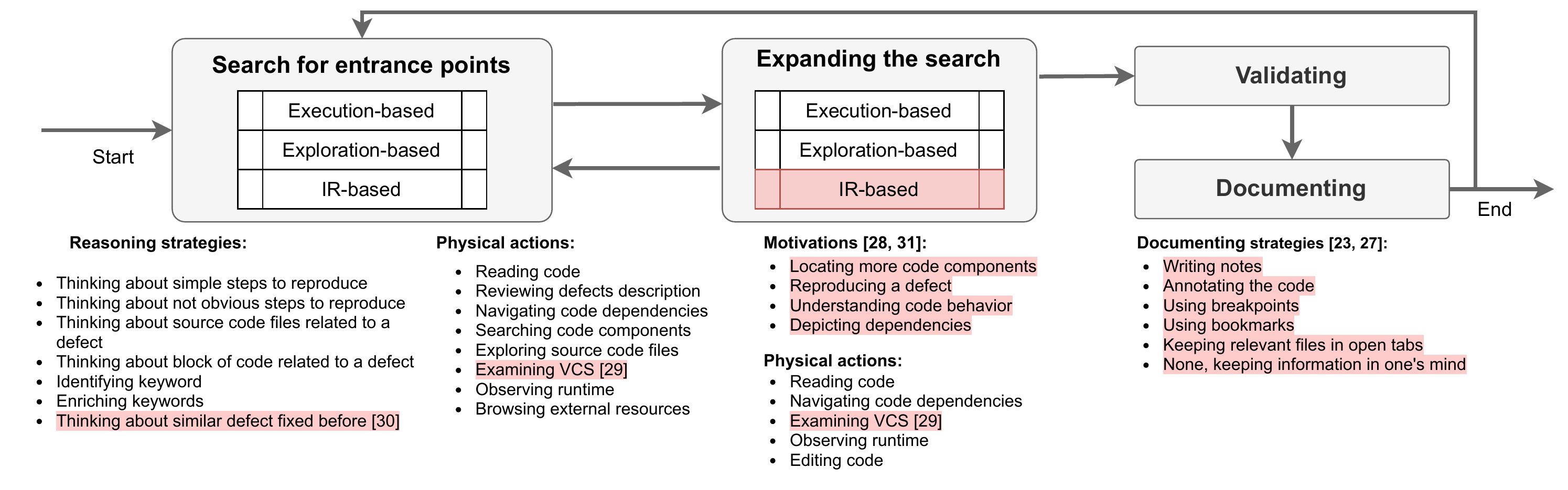}
    \caption{Defect location model based on Wang et al.~\cite{wang2013}.}
    \label{fig:defect-location-model}
     \vspace{-5mm}
\end{figure*}

We deem this model serves as a good starting point as it: (1) captures feature location process at different levels of granularity; (2) is based on three exploratory studies with a large number of developers; and (3) shares a common set of activities with previous studies.
Although the model provides a comprehensive description of patterns used for locating relevant code entities, we decided to alter some of its components to include developers' practices discovered by prior studies. These modifications capture relevant activities as indicated by the participants of the mock survey.
Figure~\ref{fig:defect-location-model} presents the overview of the model with phases containing available patterns, while physical and mental actions are listed below each phase. We mark introduced modifications with red color.
In general, we reuse all of the phases defined in the original model. However, we extend available patterns. In particular, we included the IR-based pattern in the \textit{Expand} phase for the symmetry with the \textit{Search} phase. 
Researchers noted that developers tend to lose track of the relevant program elements due to a large number of examined files and interruptions~\cite{ko2006,wang2013,sillito2008,latoza2006}. 
To provide more context for the \textit{Document} phase and investigate if/how developers preserve the information about relevant code components, we curated a list of common documenting strategies based on prior work~\cite{ko2006, wang2013}.

We modified the original list of physical actions in the following ways. We merged {\em Breakpoints operations}, {\em Step program} and {\em Run program} into one succinct action, {\em Observing runtime}. As noted in \cite{kruger2018}, the location of a specific feature can be discovered by examining the release log and pull requests, thus to reflect that we added new physical action, {\em Examine Version Control System (VCS)}. 
We included all mental actions identified by Wang et al. and added one new action, {\em Thinking about similar defects fixed before}, which aims to measure the importance of developers experience~\cite{jordan2015}. We decided to refer to mental actions as {\em reasoning strategies} throughout the survey for ease of understanding. Finally, we associated a set of {\em Motivations/Goals} with the {\em Expanding} phase to capture the main objectives of developers at this stage ~\cite{ko2007information,sillito2008}.

\subsection{Survey Design}
\label{sec:survey-design}
We leveraged the taxonomies developed in the previous steps to formulate our survey questions and ask participants to provide answers in the context of their routine work. We divided the survey into three main sections: \textit{Software Defects, Defect Location,} and \textit{Demographics}. 

In the \emph{Software Defects} section, we focused on the distribution of the defects taxonomy attributes in the context of typical and challenging defects. We primed the survey participants with the following definitions:
\begin{itemize}
    \item \emph{Typical defects} - software defects faced frequently by study participants in their daily work,
    \item \emph{Challenging defects} - software bugs that require a considerable amount of time and effort when fixing due to, e.g., difficulties in isolating the root cause, inability to reproduce the defect, or developing a viable solution.
\end{itemize}
Participants were asked to select between 3 to 5 typical software defect categories and up to 3 challenging categories. A short explanation accompanied each defect category. To capture the properties of typical and challenging bugs, we presented developers with a 5-point Likert scale (Never, Rarely, Sometimes, Often, Always), asking how frequently the participant observes specific properties of a defect, e.g., reporter being a tester or an end-user.

In the \emph{Defect Location} section, we focused on collecting the distribution of taxonomy attributes in the context of a general defect location process. We asked questions about the most useful patterns, reasoning strategies or motivations, and physical actions related to the defect location process. Moreover, to investigate how developers move through the code, at the end of each phase, we asked them about their preferred granularity level~\cite{kocchar2016}.

% In the section related to \emph{Tool Preferences}, we asked participants about their preferences of an idealized defect location tool attributes. Specifically, we asked about 
% \begin{itemize}
%     \item interaction modalities (such as visual only, chatbots, text only, and voice only)
%     \item Acceptable False Positive and False Negative rates
%     \item Ranked attributes (quality ...) 
% \end{itemize}

Finally, we captured the demographic information in the \textit{Demographics} section. In particular, we asked about:
\begin{itemize}
    \item Years of paid professional experience,
    \item Preferred programming languages,
    \item Current country of residence,
    \item Willingness to participate in a follow-up interview.
\end{itemize}

Once we had the initial draft of questions, we ran a mock survey with both COBOL and non-COBOL developers following a think-aloud protocol~\cite{redmiles2017summary} to estimate the average response time and further fine-tune our survey. The mock survey was performed with developers recruited in-house at Phase Change Software via teleconference calls. We were interested in identifying the questions that were often: (1) \textit{misunderstood} - participants consistently erred in understanding the intent behind the question; (2) \textit{irrelevant} - participants consistently question the relevance of the question.

We promptly removed the irrelevant questions. Furthermore, we did a lightweight thematic analysis of the think-aloud recordings to further refine the questions, which included three changes. First, we refined the language in questions that caused of the confusion. Second, we reordered questions to achieve a better transition. Finally, we created two distinct surveys individually catering to the COBOL and the non-COBOL developer population. The surveys had semantically equivalent questions, differing, when necessary, to account for discrepancies in terminology. For instance, classes/files in the non-COBOL survey translate to modules/procedures in the COBOL survey. 
Based on our mock sessions, we estimated that our survey’s average response time is 30 minutes. All of the questions used in our final survey can be found on the project website~\cite{projectWeb}.

\subsection{Survey Deployment and Participant Recruitment}

To recruit COBOL developers,
%from the former group
we contacted the mailing list maintained by Phase Change Software\footnote{\url{https://www.phasechange.ai/}}. We further posted the survey to various COBOL groups on Linkedin and IBM-maintained COBOL developer forums.
To recruit non-COBOL developers, we reached out to our personal contacts at several worldwide software companies (Spotify, Google, Facebook, Nokia, Capgemini), asking to distribute the survey across experienced software developers. To capture the perspective of the broader audience of software engineers, we also publicized our survey on professional forums within social platforms such as LinkedIn, Twitter, and Facebook. 

To improve our response rate, we offered rewards in the form of a chance to win an Amazon gift card. For COBOL developers, we set the reward to USD 25 for 25 participants selected at random. For non-COBOL developers, we set the reward to USD 10 for 25 participants selected at random. The disproportionate number of developers influenced the difference in reward in each group, which is also reflected in these two groups’ response-rates.

\subsection{Analysis and follow-up validation}
In this stage, we analyzed collected responses. A summary of these results was presented to a subset of survey participants as a light-weight mechanism to confirm our synthesis. 
Each semi-structured session lasted 30 minutes and followed a script prepared beforehand.
In all, we recruited 6 COBOL and 4 non-COBOL survey participants at random for the follow-up interviews. 
These sessions were conducted on Zoom, an online video conferencing platform, and recorded with the participants’ permission to be transcribed and analyzed offline.

\subsection{COBOL Forums Defect Classification}
\label{forumDefectClassification}
Studying developers' perspectives can be biased by one's perception; hence we decided to compare defect types reported in the survey with results obtained by data mining.
We referred to prior research~\cite{morrison2018} to triangulate the survey results of the non-COBOL reported defect types. However, no such baseline existed for COBOL developers. To address this gap, we manually classified the queries posted in the COBOL Programming forums of the IBM Mainframe Experts group~\cite{ibmmainframes}, which we believe is representative of our target COBOL developers population. 

The forum has over 6000 posts, so to get to a 95\% confidence level with a 10\% confidence interval to the target representative population, we needed to annotate at least 100 posts. However, not all of the posts were defect-related. Thus, authors ended up coding 400 posts only to discard nearly 300 posts as non-defects. 
Authors independently classified the defect types of 20 defect-related posts. We captured our agreement with a 0.32 Fleiss' Kappa score, which can be interpreted as a ``fair agreement''. After discussing and resolving the differences, the authors again classified the next 20 defect-related posts to reach 0.64 Fleiss' Kappa score indicating ``substantial agreement’’. The authors then resolved their differences and split the rest of the posts to code them independently. In all, we coded 107 defect-related posts.

\section{Results}

\subsection{Demographic Information}

Overall, we received 106 responses from software developers located across 11 countries, with the majority of participants residing in the United States, India, and Poland. We discarded the responses from two developers who reported having less than a year of professional industry experience and considered the remaining responses for further analysis.
Out of 104 respondents, 30 are currently, or were recently, involved in a COBOL project, while the remaining 74 work with modern programming languages, including Java, Python, and C\#. In the remainder of this section, we refer to the first group as {\em COBOL} and the later as {\em non-COBOL} developers.

Among the COBOL developers, we noted 25 males and 5 females. Non-COBOL developers included 58 males and 10 females, while 6 developers did not disclose their gender.
On average, COBOL developers reported to have 22.3 years of professional programming experience ($std=14.84$, $min=4$, $max=60$). In contrast, non-COBOL developers reported to have 9.78 years of professional programming experience ($std=7.68$, $min=1$, $max=51$).

\subsection{Software defects}
\begin{figure*}[t!]
    \centering
    \begin{subfigure}[t]{0.4\textwidth}
        \centering
        \includegraphics[width=\columnwidth]{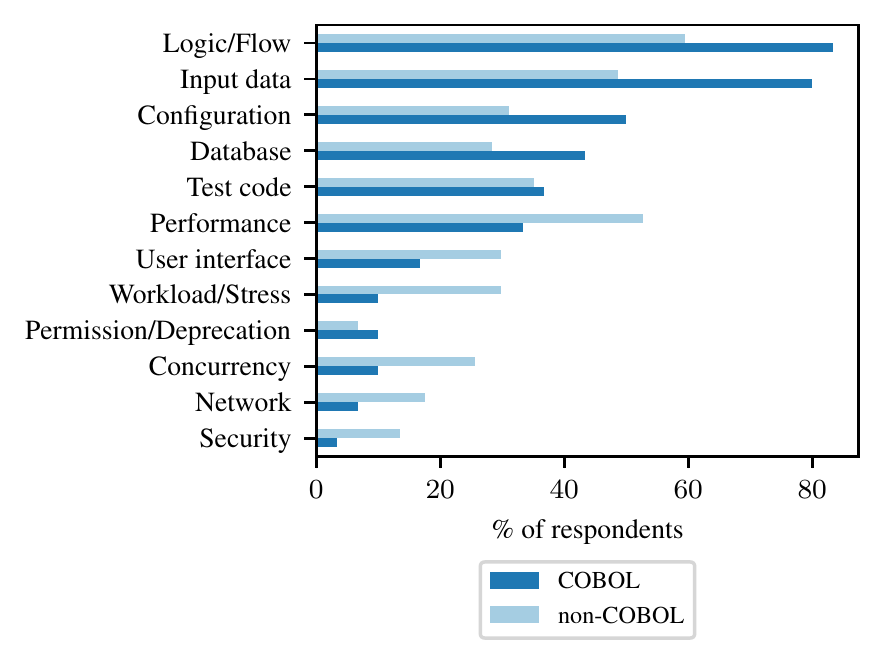}
        \caption{\footnotesize{Q: \textit{What defect categories do you work on most frequently?}}}
        \label{fig:defects-typical}
    \end{subfigure}%
    ~ 
    \begin{subfigure}[t]{0.4\textwidth}
        \centering
        \includegraphics[width=\columnwidth]{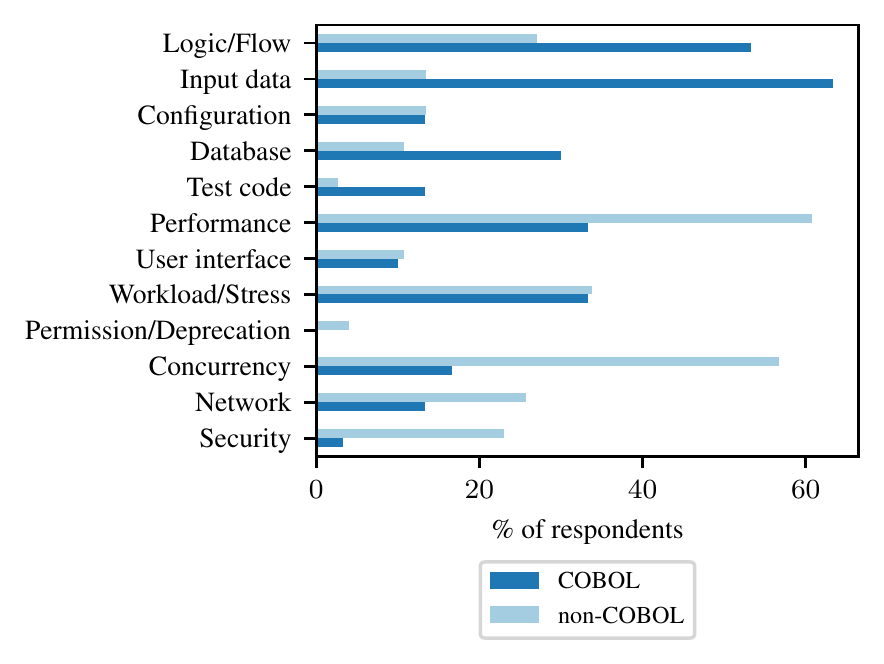}
        \caption{\footnotesize{Q: \textit{What are the most challenging defects categories?}}}
        \label{fig:defects-challenging}
    \end{subfigure}
    \caption{\small{Developers' perspective on typical (left) and challenging software defects (right).}}
    \label{fig:defects}
     \vspace{-5mm}
\end{figure*}

\noindent\textbf{Typical Software Defects.}
Figure~\ref{fig:defects-typical} shows the distribution of developers' perception of typical defect types.
Logic/Flow and Input data are the most frequently occurring defect categories according to both COBOL and non-COBOL developers. Logic/Flow was selected by 83.3\% of COBOL and 59.5\% of non-COBOL developers, while Input data was reported by 80.0\% and 48.6\% of COBOL and non-COBOL developers, respectively.
Our results are in conformance with previous work that states that 
the majority of software defects are related to programming errors leading to exceptions, crashes, or incorrect behavior~\cite{catolino2019,tan2014}.

Developer perception varies for the rest of the defect categories. COBOL developers indicated that other typical defects pertain to Configuration (50\%) and Database (42.9\%), whereas Security, Network, or Concurrency related defects rarely happen. We posit that the lack of focus on modules implementing network or security features within COBOL mainframe batch processing explains this distribution.
In contrast, non-COBOL developers reported Performance as their major category of frequently occurring typical defects (52.7\%). Overall, we notice that defects related to the rest of the categories, Configuration, Database, Test Code, User Interface, Workload/Stress, and Concurrency, are fairly evenly distributed, being reported as common by 25.7\% (Concurrency) to 35.1\% (Test Code) of non-COBOL developers.

We ran a chi-squared test to test the null hypothesis: \textit{``the distributions of defect types among COBOL and non-COBOL responses are the same’’}.  
However, low values reported in various defect categories by our COBOL developers required us to follow the common recommendation to combine some of the rows to obtain sufficient values for an approximation of a chi-square distribution~\cite{ott1988introduction}. 
We combined Network, Security, and Permission/Deprecation into a  ``Miscellaneous’’ category. 
Based on the $p$-value of 0.02, we reject the null hypothesis, concluding that typical defect types encountered by COBOL developers \textit{are} significantly different from typical defect faced by non-COBOL developers. 

We investigated whether developers' perception of typical defect types follows a distribution reported in previous studies. As a baseline for non-COBOL responses, we selected data from Morrison et al.~\cite{morrison2018} that includes a manual annotation of defect types from large open-source projects. 
Since our survey asked developers to select the top-3 typical defect types, we compared the survey results with the frequencies of top-3 defect categories reported by the baseline (Performance, Installability, and Capability). Note that we mapped categories of defects used in the baseline to defect types leveraged by this study. 
We ran a chi-square test to test the null hypothesis that ``\textit{defect categories reported by the survey and baseline follows the same distribution}''. The test resulted in $p=0.16$; thus, we accept the null hypothesis, which indicates that non-COBOL developers' perceptions of typical defects are fairly accurate.

In the case of COBOL responses, we use annotated defect-related posts from the IBM mainframe forum as a baseline (described in Section~\ref{forumDefectClassification}). We observed that the top-3 most discussed defect types are related to Logic/Flow, Configuration, and Input data, which broadly confirms the survey results.

\begin{tcolorbox}
\small{\textit{RQ1}: Defects related to Logic/Flow and Input data are the primary defect types in COBOL systems. Typical defect types encountered by COBOL developers are significantly different from typical defects encountered by non-COBOL developers.}
\end{tcolorbox}

% challenging defects
\noindent\textbf{Challenging software defects.}
The distribution of developers' perception of challenging defect types is shown in Fig.~\ref{fig:defects-challenging}.

According to COBOL developers, challenging defects are mostly related to Input data (64.3\%) and Logic/Flow (53.6\%), making those categories typical and challenging simultaneously. As the top third most challenging defect 35.7\% of COBOL developers selected Workload/Stress. In contrast, only 12.5\% of COBOL developers reported Workload/Stress as a \textit{typical} defect.
Non-COBOL developers indicated issues related to Performance (60.8\%), Concurrency (56.8\%), and Workload/Stress (33.8\%) as the 
most challenging, whereas Logic/Flow and Input data defects were selected by 27\% and 13.5\% of the respondents, respectively.

We ran the chi-squared test to compare the distribution of challenging defect types among COBOL and non-COBOL developer responses. In particular, we tested the following null hypothesis:  \textit{``challenging defect types reported by COBOL and non-COBOL developer responses follow the same distribution}’’. Our test resulted in $p < 0.01$. We therefore reject the null hypothesis, indicating that challenging defects reported by COBOL and non-COBOL developers \textit{are}, in fact, different.
We also performed the chi-squared test to evaluate the differences in the distribution of typical and challenging defects within each respondent group. Specifically, for each group, we tested the null hypothesis \textit{``if typical and challenging defect types have the same distribution}’’. In the case of COBOL, our test resulted in $p=0.096$, therefore we accept the null hypothesis and conclude that typical and challenging COBOL defects are not significantly different.
In contrast, challenging defects reported by non-COBOL respondents are significantly different from the typical defect types with $p\ll0.01$.

In the follow-up interviews, most COBOL and non-COBOL developers agreed with the top typical and challenging defect categories reported by their respective groups. COBOL developers indicated that Logic/Flow and Input data could be challenging since \textit{there is no good IDE for COBOL compared to IntelliJ or Eclipse tools, and the structure of COBOL code can easily lead to programming errors}. %To support this statement, 
In support, another COBOL developer mentioned one of the hardest defects he faced was related to a missing \textit{null check for file pointer}, that could be easily detected by a modern IDE.
Developers also stated that, since \textit{COBOL is primarily a legacy system, they rarely know the entire process}, thus making it difficult to locate the root cause effectively. Finally, developers also noted that Workload/Stress and Performance defects pose a challenge due to the difficulty in local reproduction and the unique characteristics of every defect.
Overall, developers listed the following challenges associated with specific defects: (1) \textit{reproducing the defect outside the production environment is often impossible}; (2) \textit{defects do not surface immediately; instead, they happen occasionally, leading to numerous problems across components}; %and, finally, 
(3) \textit{it can take days to find the right solution}.

\begin{tcolorbox}
\small{\textit{RQ2:} Challenging defects encountered by COBOL developers are not different from typical defects. However, challenging defects encountered by non-COBOL developers vary significantly from typical defects.}
\end{tcolorbox}

\begin{figure*}[ht!]
    \begin{subfigure}{0.40\textwidth}
        \centering
        \includegraphics[scale=0.7]{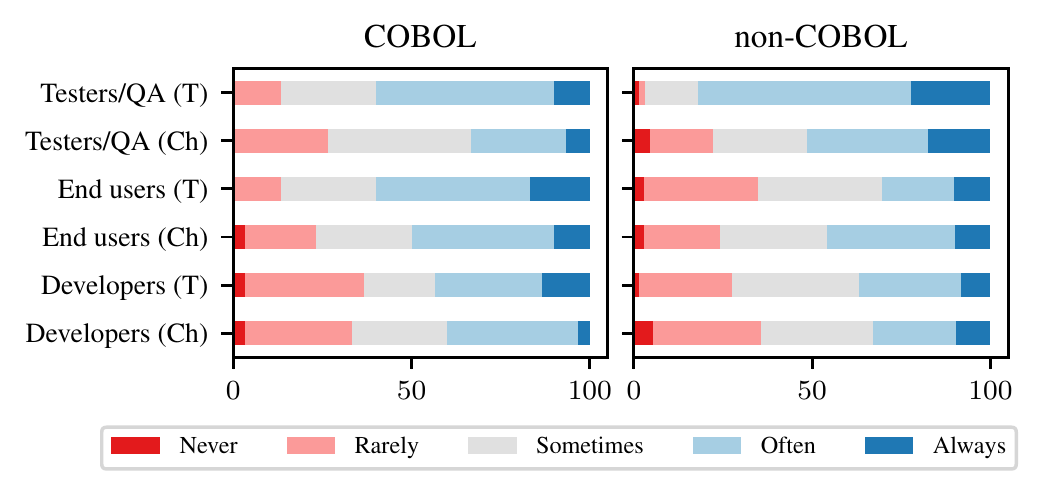}
        \caption{\footnotesize{Q: \textit{How frequently are software defects reported by:}}}
        \label{fig:reporter}
    \end{subfigure}%
    ~
    \begin{subfigure}{0.55\textwidth}
        \centering
        \includegraphics[scale=0.7]{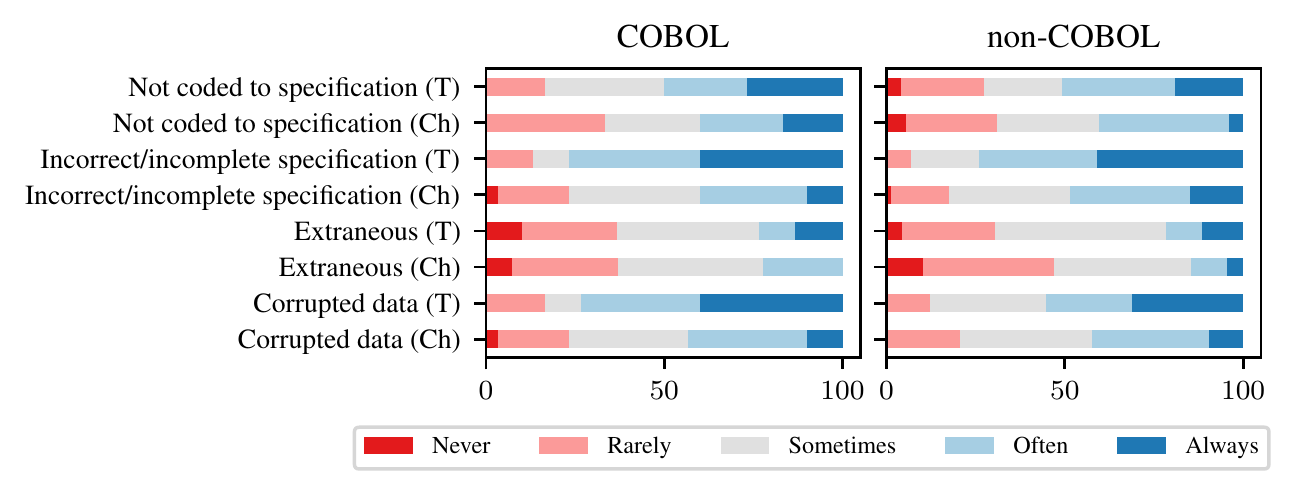}
        \caption{\footnotesize{Q: \textit{How likely is a software defect caused by:}}}
        \label{fig:cause}
    \end{subfigure}

    \begin{subfigure}{0.45\textwidth}
        \centering
        \includegraphics[scale=0.7]{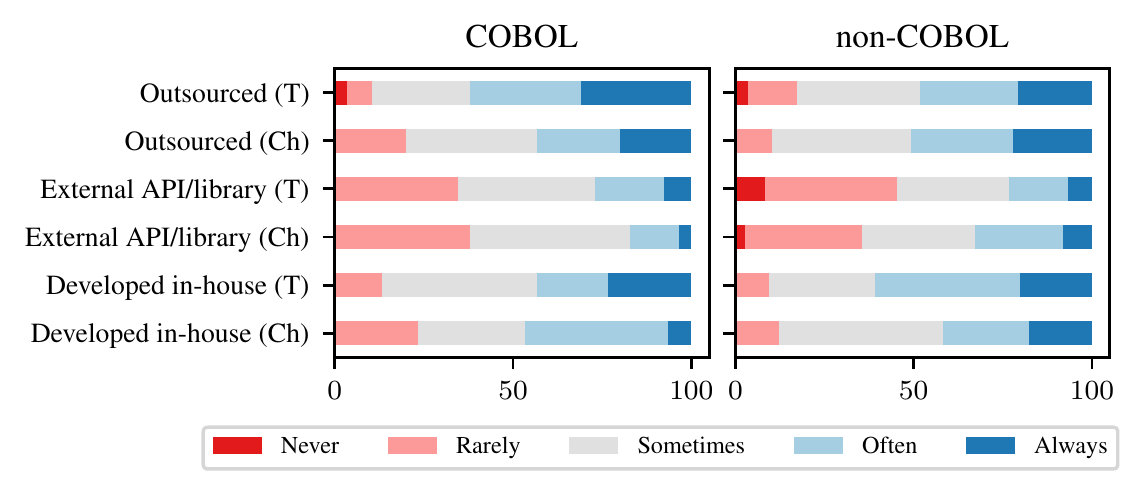}
        \caption{\footnotesize{Q: \textit{How likely is a software defect to occur in code that is:}}}
        \label{fig:source}
    \end{subfigure}\hspace{5mm}%
    ~
    \begin{subfigure}{0.50\textwidth}
        \centering
        \includegraphics[scale=0.7]{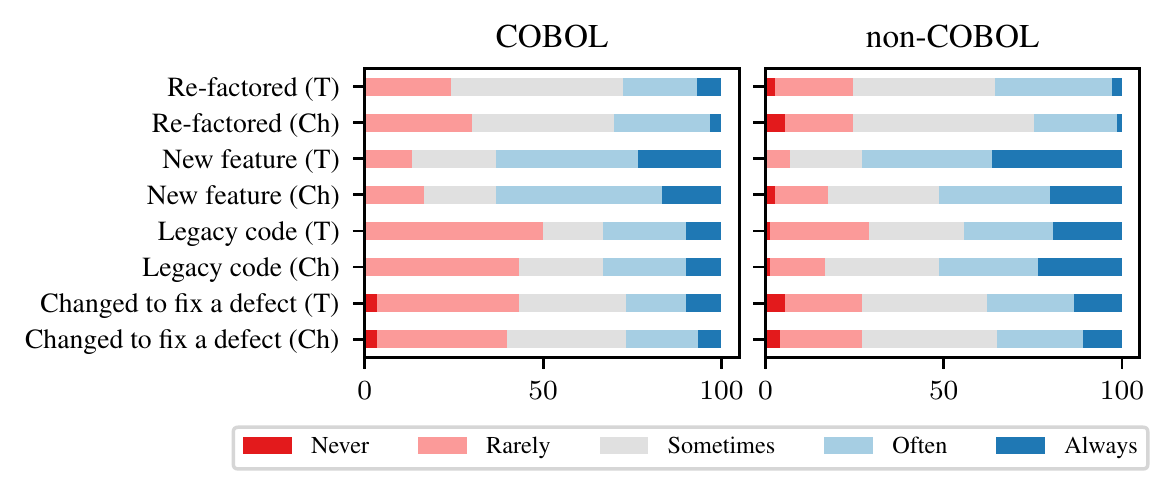}
        \caption{\footnotesize{Q: \textit{How likely is a software defect to occur in code that is:}}}
        \label{fig:age}
    \end{subfigure}%
    \caption{\small{Developers' perspective on the properties of software defects.}}
    \label{fig:properties}
    \vspace{-3mm}
\end{figure*}

\noindent\textbf{Defect properties.}
Developers' perspectives on the properties of software defects are illustrated in Fig.~\ref{fig:properties}, with each property visualized as a Likert scale. Properties of typical defects are marked with (T), whereas challenging defects with (Ch).

When asked about defect reporters, 63.2\% of COBOL and 80.7\% of non-COBOL developers agreed that testers are the most likely group to detect typical defects (Fig.~\ref{fig:reporter}). %On the contrary, 
In contrast, only 33.3\% of COBOL and 47.3\% of non-COBOL developers expressed positive opinions about testers’ ability to detect challenging defects.
At the same time, both groups indicated that end-users are more likely to report challenging defects, with 50\% for COBOL and 43.2\% for non-COBOL supporting such opinion.

We conducted unpaired Mann-Whitney tests to verify the null hypothesis: \textit{``testers have the same capability to detect typical and challenging defects’’}. The tests were conducted for COBOL and non-COBOL separately. We note $p=0.029$ and $p\ll0.01$ for COBOL and non-COBOL, respectively, indicating a statistically significant difference in perceived testers’ abilities. We conducted analogous tests for end-users and obtained $p=0.052$ and $p=0.143$; thus, we conclude that the observed difference is not significant.

In the follow-up interviews, both COBOL and non-COBOL developers supported these findings indicating that \textit{since end users have a system that has already been tested, then all the easy defects have been already caught. Hence, end-users usually find more complicated issues, which the engineering team has not thought of}. Moreover, testing for, %challenging defects, 
e.g., Performance or Workload/Stress, \textit{is rarely doable in the development environment} and \textit{requires significant time and effort}.

COBOL and non-COBOL developers agree that typical defects are mostly caused by corrupted data (73.3\% and 55.4\%) or missing specifications (76.7\% and 73\%). 
In the case of challenging defects, respondents expressed weaker opinions about the root cause, shifting their opinions from \textit{Always} towards \textit{Often} and \textit{Sometimes}. Additionally, we note an increased selection of \textit{Not coded to specification} as a root cause of challenging defects compared to the results obtained for typical defects. 
Developers also agree that both typical and challenging defects are more likely to be located in the code developed in-house or outsourced than in %the
external APIs (Fig.~\ref{fig:source}). 
The majority of defects are deemed to originate in a newly developed code. However, developers ranked legacy code as the second most likely source of issues (Fig.~\ref{fig:age}). 

\begin{tcolorbox}
\small{Testers' abilities to detect challenging defects are limited due to environments’ restrictions and lack of tools supporting efficient testing of e.g., Performance, or Workload/Stress-related scenarios.}
\end{tcolorbox}

\subsection{Defect Location Strategies}

We analyzed developers' responses for the three phases of the defect location process outlined in the taxonomy in Section~\ref{studyDesign_defectLocationTaxonomy}: searching for a starting point, expanding relevant program elements, and documenting~\cite{wang2011}.

\noindent \textbf{Phase 1: Search.}
Table~\ref{tab:start-mental} presents reasoning strategies employed by developers in the first phase of the defect location process, ranked in order of importance. We note that each group of respondents have their own distinctive preferences. COBOL developers declare to first focus their efforts on locating a similar defect that was solved before (47\%), while non-COBOL developers prefer to identify simple steps to reproduce (44\%). 
Moreover, both groups frequently selected the opposite strategy as their second top choice, further emphasizing the usefulness of the two approaches. Next, participants indicated they tend to look into source code files and blocks of code, and finally, they concentrate on keywords present in a defect report. 

The developers further confirmed these rankings in the follow-up interviews. For instance, COBOL developers mentioned that \textit{seeing a defect which is similar to an issue fixed before gives a good approximation of the potential problem and its location}, therefore their organizations tend to \textit{keep defect databases to retain the knowledge of previous patterns of software failures}.

\begin{table}[t]
\scriptsize
\caption{Reasoning strategies undertaken by developers when examining new defect reports.}
\label{tab:start-mental}
\centering
\scriptsize
\begin{tabular}{p{4.8cm}|ll}
\toprule
\multirow{2}{4.8cm}{Reasoning strategy} & \multicolumn{1}{l}{COBOL} & \multicolumn{1}{c}{Non-COBOL} \\
 & Mean rank & Mean rank  \\
\midrule
Thinking about similar defect fixed before & 2.567 (1) & 3.378 (2) \\
Thinking about simple steps allowing to reproduce the defect & 3.133 (2) & 2.568 (1) \\
Thinking about source code files that may be related to the defect & 3.200 (3) & 3.689 (4) \\
Thinking about block of codes that may be related to the defect & 4.167 (4) & 3.500 (3) \\
Identifying keywords that can be related to defect's location & 4.367 (5) & 4.014 (5) \\
Refining or augmenting keywords based on my knowledge & 5.567 (6) & 5.541 (6)\\
Thinking about not obvious steps that can cause the defect & 5.000 (7) & 5.311 (7)\\
\bottomrule
\end{tabular}
\end{table} 

\begin{figure}
    \centering
    \includegraphics[width=\columnwidth]{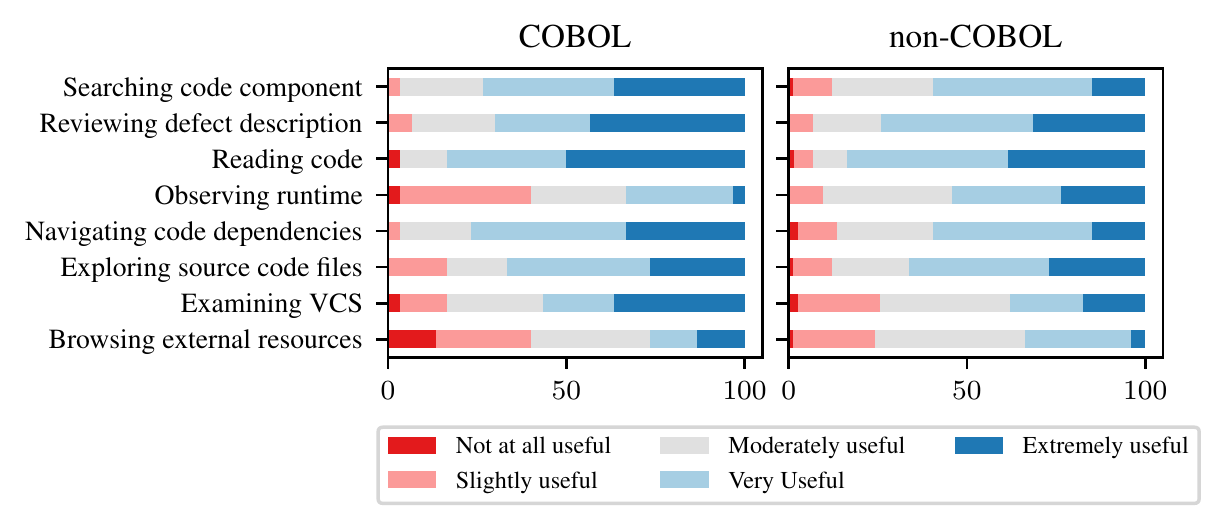}
    \caption{\footnotesize{Developers' perspective on the usefulness of physical actions in the Search phase.}}
    \label{fig:start-actions}
     \vspace{-5mm}
\end{figure}

Developers' opinion on the effectiveness of physical actions is illustrated in Fig.~\ref{fig:start-actions}. In the initial phase, developers rank reading code and reviewing defect descriptions as the most useful actions for locating an entry-point for further exploration. Additionally, we note that COBOL developers prioritized searching for code components and navigating code dependencies, whereas non-COBOL developers prefer to explore source code files.
This difference can be attributed to the tool capabilities (e.g., providing a summary of related code components) used by non-COBOL developers during their quick file exploration.

With respect to the most useful patterns in the Search phase of the defect location process, the majority of developers in both groups indicated they prefer to use an execution-based strategy to locate an initial set of entry points, followed by textual search and navigating static code dependencies. 

\noindent \textbf{Phase 2: Expand.}
After locating a set of entry points, developers move to the Expand phase. Table~\ref{tab:expand-mental} depicts the main goals developers aim to achieve at this stage. We observe that both groups are mainly focused on reproducing the defect to confirm whether their location process moves in the direction of the defect's cause. Additionally, respondents indicated that they need to understand the code behavior better. Follow-up interviews revealed that as developers are often under time pressure, they prefer to focus on the fastest path leading to resolving the defect. Therefore, they opt for reproducing the faulty behavior as \textit{once the defect gets reproduced, it is very easy to proceed further}.

\begin{table}[h]
\scriptsize
\caption{\small{Developers' motivations and goals in the Expand phase.}}
\label{tab:expand-mental}
\centering
\scriptsize
\begin{tabular}{p{4.8cm}|ll}
\toprule
\multirow{2}{4.8cm}{Motivation/Goal} & \multicolumn{1}{l}{COBOL} & \multicolumn{1}{c}{Non-COBOL} \\
 & Mean rank & Mean rank  \\
\midrule
Reproducing a defect & 1.600 (1) & 1.905 (1) \\
Understanding code behavior & 2.500 (2) & 2.162 (2) \\
Locating more code components & 2.633 (3) & 2.824 (3) \\
Depicting dependencies & 3.267 (4) & 3.108 (4) \\
\bottomrule
\end{tabular}
\end{table}

\begin{figure}
    \centering
    \includegraphics[width=\columnwidth]{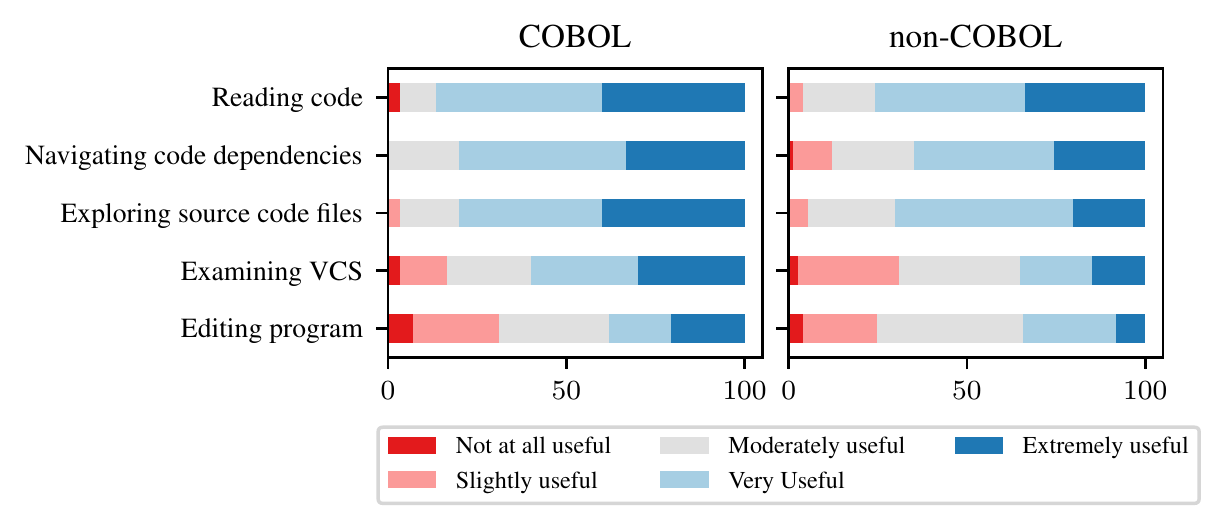}
    \caption{\footnotesize{Developers' perspective on the usefulness of physical actions in the Expand phase.}}
    \label{fig:expand-actions}
    \vspace{-5mm}
\end{figure}

Developers' perception of the usefulness of physical actions in the Expand phase is shown in Fig.~\ref{fig:expand-actions}. Like the Search phase, developers rank reading code as the most beneficial activity, followed by navigating structural code dependencies and exploring source code files.
Further, when analyzing developers’ patterns in this phase, we note that they prefer to use an execution-based approach. In contrast to the Search phase, developers reported leveraging code dependencies more often than performing a textual search.

\begin{figure}
    \centering
    \includegraphics[width=0.75\columnwidth]{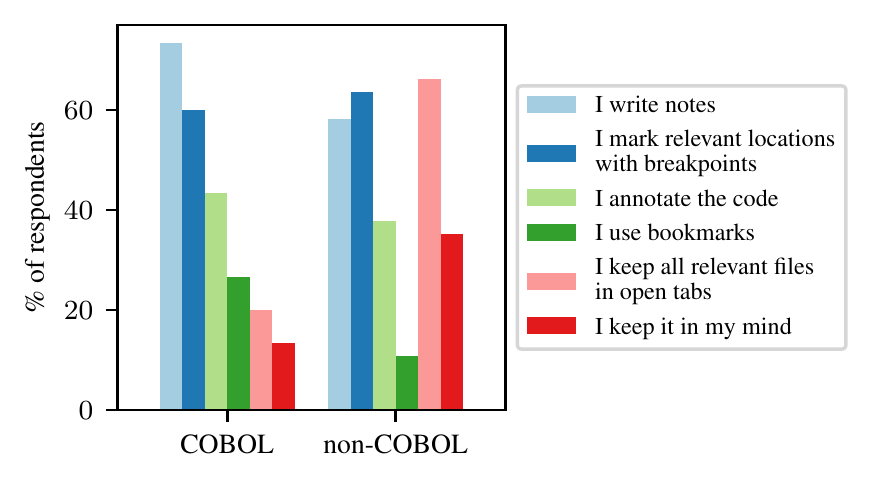}
    \caption{\footnotesize{Developers' perspective on documentation.}}%the document phase}
    \label{fig:document}
     \vspace{-5mm}
\end{figure}

\noindent \textbf{Phase 3: Document.}
Preserving information about located relevant files concludes the defect location process. Fig.~\ref{fig:document} illustrates how developers keep track of important code components. Respondents in both groups indicated they frequently write notes (e.g., on paper or in a text editor) and use breakpoints to indicate relevant code pieces. We notice that non-COBOL developers rely more on ``open tabs’’ and ``keep in mind’’ strategies compared to COBOL developers. Both preferences can be explained in part by the tooling support for non-COBOL developers. For instance, a typical modern IDE provides ample opportunity to keep multiple tabs open and enough visual cues to facilitate the ``keep in mind’’ strategy.

\begin{tcolorbox}
\small{\textit{RQ3:} COBOL and non-COBOL developers approach defect location rather similarly. Minor differences are explainable by the varying level of experience and available tooling support.}
\end{tcolorbox}

\subsection{Code granularity throughout the defect location process}

In the case of non-COBOL developers, we observe a consistent preference for Method/Function level code granularity across all phases with support varying between 37.8 and 44.6\%. We note a slight change in preferences in Phase 3 when developers mark relevant code components to be fixed. At this stage, non-COBOL developers reported increased interest in the Block of code (from 17.6\% to 30\%) and Line of code (from 4.1\% to 21.6\%) levels of granularity.
COBOL developers tend to follow a top-down approach focusing first on the high granularity level and progressively going towards lower levels. In the Search phase, 50\% of COBOL developers prefer to work with Files or Methods. In the Expand phase, they increasingly prefer Methods level, while when documenting, they operate at the level of Blocks or Lines of code.

All COBOL developers agreed with these results during follow-up interviews and related the top-down code navigation hypothesis to the COBOL code structure.
On the other hand, non-COBOL developers indicated that while they sometimes follow a similar top-down approach, overall, they prefer to 
reason about functions as \textit{a single line of code rarely provides enough context}, whereas, in \textit{a well-written code, a method encapsulates a specific sub-task and provides just enough information to understand and efficiently solve the defect}.

\begin{tcolorbox}
\small{COBOL developers typically follow a top-down approach to navigate artifacts during defect location as reflected by their choices for specific artifacts at different phases.}
\end{tcolorbox}

\begin{figure}
    \centering
    \includegraphics[width=\columnwidth]{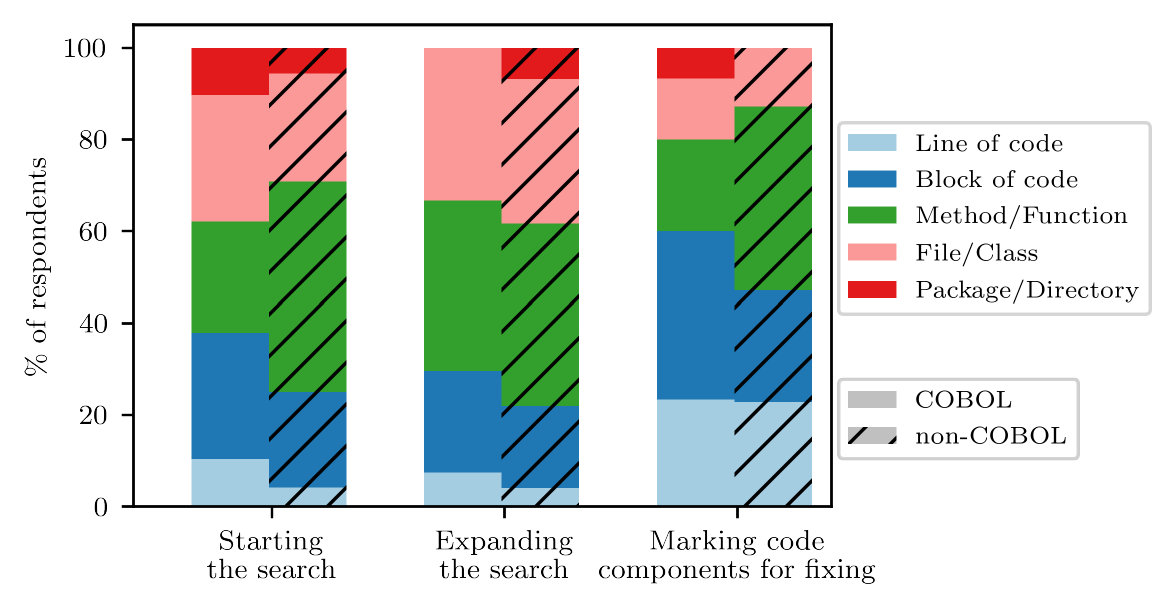}
    \caption{\footnotesize{Preferred granularity of code components in different phases of defect location process.}}
    \label{fig:granularity}
     \vspace{-5mm}
\end{figure}

\section{Discussion and Future Work}

\noindent
\textbf{Typical defects are challenging.}
We observed that there is no significant difference between typical and challenging defects in COBOL.
In general, COBOL developers reported Logic/Flow and Input data as top-two major defect categories, which are both typical and challenging. This indicates that a hypothetical defect location tool addressing just these two defect types is likely to significantly impact the developer productivity and reduce the time required to locate and fix the defect. As indicated by a COBOL developer in a follow-up interview, the lack of tools targeting COBOL is one of the key factors contributing to Logic/Flow being a challenging defect. On the contrary, non-COBOL developers, who enjoy an wide set of tools, reported that typical defects such as Logic/Flow or Input Data rarely pose any difficulties. We believe that if COBOL developers are given supporting tools addressing their most pressing needs, we would observe a decrease in Logic/Flow and Input data defects’ perceived difficulty.

 COBOL and non-COBOL developers reported that challenging defects are often not detected by testers; hence they reach production. While undesirable in any application, it is even worse for COBOL ones as they typically support critical financial, health care, and government systems. Our respondents stated that the key reason for challenging defect propagation is related to differences between testing and production environments’ configuration and capabilities. Taking a cue from the continuous delivery movement~\cite{humble2010continuous}, we recommend tool-smiths to cost-effectively minimize the differences between COBOL development and production environments. 

\noindent
\textbf{Defect location strategies are fairly similar.} 
Even though our respondents work with different software ecosystems, they reported following fairly similar defect location strategies, indicating that observations and findings made by prior studies in the context of common patterns followed by non-COBOL developers can be applied to COBOL developers. %as well. 
Although confirmation requires further research, initial results are encouraging. Analyzing the survey results, we also note the benefits that a simple code navigation tool~\cite{henley2018human} could provide to COBOL developers. First, the tool could help developers keep track of the code under analysis as indicated in the code documentation preferences. Further, it could also be aligned with observed top-down navigation patterns where COBOL developers start at the highest granularity (File/Module) and iteratively move to finer granularities (Method/Paragraph).

%We also make a note of COBOL developers could really benefit from a simple code navigation tool~\cite{henley2018human} to help keep the code under analysis in the context as indicated by the choices selected in code documentation preferences. Furthermore, an effective tool for code navigation should also align with observed top-down navigation patterns where COBOL developers start with the highest granularity level (File/Module) and iteratively move to a finer granularity (Method/Paragraph). 

Respondents in both groups indicated the Execution-based pattern, involving running or debugging an application, as the most effective strategy for defect location. However, whereas non-COBOL developers usually have unrestricted access to the application via an IDE, the same cannot be said for COBOL developers. Mainframe time is expensive, so the possibility of debugging or re-running the application is often limited. On the other hand, follow-up interviews revealed that the sooner developers could reproduce a defect, the faster they can resolve it. Thus improving the state-of-the-art of debugging in the mainframe environment would greatly benefit developers.

%\noindent\textbf{COBOL and non-COBOL developers exhibit different navigation patterns.} While non-COBOL developers prefer to focus on the Method/Function level when locating a defect, COBOL developers follow a top-down approach. This result is somehow intuitive, as the top-down navigation through code reflects the COBOL code structure. Moreover, as stated by a COBOL developer, when a defect occurs, the first information a developer receives pertain to the highest granularity level, therefore the developer will typically start investigation there and move towards lower granularity progressively. This result implies that a tool supporting COBOL developers may need to operate at various levels of granularity to be actually useful.

\noindent
\textbf{Emerging research directions.}
For the researchers, we recommend expanding the content and scope of this study. One possible research avenue is to conduct an in-situ study of COBOL developers to verify and refine actionable insights towards tools supporting mainframe development. We suspect that such an undertaking would be very fruitful as an industry collaboration since COBOL projects are typically highly regulated and not accessible to a broader audience. Another direction can be to offer COBOL courses at universities and study the student population to get end-user perspectives~\cite{ko2004six} on tool usage~\cite{pandita2018no}.

While studying developers' perspectives on software defects gives a good initial overview of defects in the COBOL ecosystem, analysis of actual COBOL defects is imperative to validate our findings~\cite{devanbu2016belief}. Our data is publicly available at~\cite{projectWeb}. We hope researchers and practitioners will curate additional benchmarks to foster future research. We concede that collecting such information is challenging since enterprise COBOL projects are either proprietary and regulated.% or open-source student toy examples.

Finally, we plan to follow previous studies~\cite{morrison2016veteran,jordan2015} and work with veteran developers to identify optimal strategies to preserve their knowledge and investigate how to remove entry-level barriers by passing hard-earned expertise onto a new generation of developers. With one of the participants reporting 60 years of programming experience, we deem this research direction to be particularly exciting and viable in the context of COBOL.

\section{Threats to validity}
This study suffers from several limitations that can impact the validity and generalizability of the results. %We discuss these and describe the steps we took to mitigate them.

\noindent\textbf{Internal validity.} Terminology employed in the mainframe environment differs from that leveraged by non-COBOL developers. Providing developers with imprecise or misleading wording may increase the interpretation bias, which, in turn, leads to unreliable results.
We developed two independent surveys with the same content to mitigate this threat, with terminology adjusted to the target group of respondents.
Examples of such adjustments include, e.g., using code granularity levels from the mainframe environment such as Job, Section, Paragraph, or modifying {\em External libraries/API} to {External libraries/COTS} (Common of the shelf software).

\noindent\textbf{External validity.} The survey measures developers' perception, which is strictly subjective and affected by interpretation bias; hence it may not reflect true defect and defect location strategies. We mitigated this threat with the following steps. First, we conducted a pilot study with a COBOL and non-COBOL developer to ensure the usage of correct and clear terminology. Second, we recruited a large number of participating developers to account for potential noise in the data. Finally, we conducted follow-up interviews to validate our observations and gather developers' rationale for them. Furthermore, we triangulated the non-COBOL defect-types results with an independent baseline~\cite{morrison2018} and COBOL defect types by mining defects from online forums. 

\section{Related work}

We contrast our work with prior research, which we grouped into three main areas: (1) COBOL, (2) developers' perception, and (3) software defects and defect location strategies.

\noindent\textbf{COBOL ecosystem.}
Despite its massive presence in the financial sector, the COBOL ecosystem has seen little attention in recent \textit{software engineering} research.
Sellink et al.~\cite{sellink2002restructuring} proposed a set of methods to effectively restructure COBOL code, while Sneed et al.~\cite{sneed2001} described the process of extracting knowledge from a legacy COBOL system.
More recently, researchers' efforts concentrated on the migration of COBOL code to Java. Mossienko~\cite{mossienko2003} introduced an automated technique translating COBOL to Java. Sneed et al.~\cite{sneed2011migrating, sneed2013migrating} reported on two industrial COBOL migration projects involving code restructuring and automated language translation.
De Marco et al.~\cite{demarco2018cobol} described migration methodology designed for a newspaper company, while Rodriguez et al. ~\cite{rodriguez2013bottom} discussed two migration approaches based on a case study of 32 COBOL transactions.
Although these studies made a significant step towards restructuring and migrating COBOL code, the ways to natively support large-scale production COBOL systems have not yet been investigated.

As any programming language, COBOL is not defect-free; however, defects and their properties in the mainframe environment were rarely investigated. Litecky et al.~\cite{litecky1976} studied errors commonly made by students learning COBOL and observed that 20\% of error types account for 80\% of errors.
Veerman et al.~\cite{veerman2006cobol} focused specifically on the \texttt{perform} statement and identified programming constructs leading to unexpected system crashes. 
%The authors argued that potentially faulty code should be inspected by a human expert due to differences in COBOL semantics.
In contrast, this study focuses on studying general defects types and defect location strategies employed by professional COBOL developers to deepen the understanding of the problems encountered in the field.

%Despite its massive presence in the financial sector, the COBOL ecosystem has not seen much attention in recent \textit{software engineering} research.
%Litecky et al.~\cite{litecky1976} studied errors commonly made by students learning COBOL and observed that 20\% of error types account for 80\% of all errors.
%Sellink et al.~\cite{sellink2002restructuring} proposed a set of methods allowing practitioners to effectively restructure COBOL code, while Mossienko~\cite{mossienko2003} introduced an automated technique translating COBOL code to Java.
%More recently, Sneed et al.~\cite{sneed2001,sneed2010} presented two industrial reports, first describing the process of extracting knowledge from a legacy COBOL system, followed by migration to Java based on the automated language translation. 
%In contrast, this study focuses on studying the defects and defect location strategies employed by professional COBOL developers to deepen understanding of the problems faced by such professionals.

\noindent\textbf{Studying developers' perceptions.}
This work leverages online surveys, a common tool employed by researchers to study developers’ expectations and perceptions.
For instance, the survey methodology has been applied to investigate debugging tools and approaches to solve the most difficult defects~\cite{siegmud2014}, and developers' expectations towards fault localization~\cite{kocchar2016}. While our work is similar in spirit, we explicitly study professional COBOL developers.
Devanbu and colleagues~\cite{devanbu2016belief} indicate that although developers’ beliefs may differ from empirical evidence, developers’ perceptions should be taken into consideration, especially when prior research is limited, and obtaining data at-scale is challenging. Although we were not explicitly investigating such discrepancies, we did find outliers in developers' perceptions. However, they were few and far between and were discarded in aggregate analysis. 

\noindent\textbf{Software defects and defect location strategies.}
Researchers and practitioners devised several classification taxonomies to capture and analyze the characteristics of software defects. Chillarege et al.~\cite{odc} introduced Orthogonal Defect Classification (ODC) comprising 8 dimensions describing defect properties. Researchers at Hewlett-Packard~\cite{grady1992} proposed a taxonomy leveraging three dimensions to explain defects' type, source, and root cause. Defect categories and properties identified in these studies have been successfully applied in various industrial settings~\cite{lutz2004,zheng2006} and led to building automatic tools for defect prevention~\cite{shenvi2009defect} and classification~\cite{huang2015autoodc}.
Recently, Catolino et al.~\cite{catolino2019} inspected defect reports from 3 large software projects to build a modern taxonomy consisting of 9 defect categories. Our work builds upon defect taxonomies proposed by Catolino et al.~\cite{catolino2019} and ODC~\cite{odc}.

Researchers conducted numerous lab and field studies observing strategies leveraged by developers while completing maintenance tasks to delineate the defect location process.
For instance, Ko et al.~\cite{ko2006} identified three main activities frequently interleaved by developers: seeking, relating, and collecting information. 
Wang et al.~\cite{wang2013, wang2011} created a hierarchical model describing the feature location process in terms of phases, patterns, and actions. Based on exploratory studies, Kevic et al.~\cite{kevic2017} identified a set of 6 re-occurring activities related to, e.g., understanding or changing source code. 
Recently, Chattopadhyay et al.~\cite{chattopadhyay2019} performed an exploratory study with 10 developers to observe how developers maintain the current task context. This work leverages the model of feature location proposed by Wang et al.~\cite{wang2013} to create the survey.

Building automatic tools to reduce the effort required to localize relevant code components has been of great interest. %Researchers proposed various techniques based on IR~\cite{corley2018,wen2016,nguyen2011topic}, static~\cite{moreno2014,wong2014}, and dynamic analysis~\cite{dao2017,bohnet2008,edwards2006approach}, leveraging different dimensions of software-related data, such as repository history~\cite{corley2018, kruger2018} and API documentation~\cite{ye2014,lam2017}. 
However, despite the plethora of advanced automated techniques for defect location~\cite{corley2018,wen2016,moreno2014,dao2017, kruger2018, lam2017}, none of these were evaluated in the context of mainframe development. Therefore, their applicability in the mainframe environment remains largely unknown.

\section{Conclusion}
This work compares the COBOL and non-COBOL developers' perspectives on software defects and defect-location strategies. We observed that: (1) defects affecting COBOL belong to two major categories(Logic/Flow and Input Data), (2) COBOL and non-COBOL developers follow similar defect location strategies, and (3) adapting debugging tools to the mainframe environment could have a significant impact in COBOL developer productivity.

\section*{Acknowledgment}
We thank our participants for their time and effort, Dan Acheff for his insights about COBOL ecosystem, and Brad Cleavenger and Greg Brueggeman for their help in designing the survey.
We also thank Phase Change Software for supporting this work.

\bibliographystyle{IEEEtran}
\bibliography{sample-base.bib}
\end{document}